\documentclass[
    aps,prc,             %
    nofootinbib,         %
    superscriptaddress,  %
    reprint,             %
    showkeys,            %
]{revtex4-2}
\usepackage{graphicx}                                 %
\usepackage{xcolor}                                   %
\usepackage{array,dcolumn,longtable}                  %
\usepackage{amsmath,amssymb,mathtools,slashed}        %
\usepackage{textcomp}                                 %

\usepackage[utf8]{inputenc}                           %
\usepackage{bm}                                       %

\providecommand\lfstyle{}                             %
\providecommand\romanup[1]{\text{#1}}                 %
\providecommand\greekup[1]{#1}                        %
\usepackage{siunitx}                                  %
\usepackage{hyperref}                                  %
\hypersetup{
    hidelinks,  %
    breaklinks, %
    unicode     %
}

\usepackage[noabbrev]{cleveref}                       %

\creflabelformat{equation}{#2#1#3}

\newcommand\CSC{\textsc{csc}}
\newcommand\doe{\textsc{doe}}

\newcommand\moe{\textsc{moe}}

\newcommand\nsf{\textsc{nsf}}
\newcommand\nsfc{\textsc{nsfc}}

\newcommand\alice{\textsc{alice}}

\newcommand\cms{\textsc{cms}}
\newcommand\phenix{\textsc{phenix}}

\newcommand\STAR{\textsc{star}}

\newcommand\clvisc{\textsc{clv}isc}
\newcommand\dabmod{\textsc{dabm}od}
\newcommand\ipglasma{IP-Glasma}
\newcommand\jetscape{\textsc{jetscape}}

\newcommand\pythia{\textsc{pythia}}

\newcommand\trento{\textsc{t\raisebox{-0.5ex}{r}ent}o}

\newcommand\eos{\textsc{e}o\textsc{s}}
\newcommand\eosPCE{s95\_\textsc{pce}}
\newcommand\qcd{\textsc{qcd}}

\newcommand\qgp{\textsc{qgp}}

\renewcommand\vec[1]{\ensuremath{\bm{#1}}}

\let\abs\undefined\DeclarePairedDelimiter\abs{\lvert}{\rvert}

\newcommand\dd{\mathop{}\!\romanup{d}}

\DeclareSIUnit{\fm}{\femto\metre}

\newcommand\proton{{\romanup{p}}}

\newcommand\pion{{\greekup{\pi}}}
\newcommand\pionzero{{\pion^0}}

\newcommand\Bmeson{{\romanup{B}}}

\newcommand\Dmeson{{\romanup{D}}}
\newcommand\Dzero{{\Dmeson^0}}

\newcommand\A{{\romanup{A}}}

\newcommand\nAu{{\romanup{Au}}}
\newcommand\nPb{{\romanup{Pb}}}

\newcommand\nuclei{{\A\A}}
\newcommand\pp{{\proton\proton}}

\newcommand\AuAu{{\nAu\text{+}\nAu}}
\newcommand\PbPb{{\nPb\text{+}\nPb}}

\newcommand\snn{\sqrt{s_\text{NN}}}
\newcommand\snnG[1]{\snn = \SI{#1}{\GeV}}
\newcommand\snnT[1]{\snn = \SI{#1}{\TeV}}

\newcommand\pt{p_\text{T}}

\newcommand\raa{R_\text{AA}}
\newcommand\vn[1]{v_{#1}}

\newcommand\Tfo{T_\text{FO}}

\newcommand\Td{T_\text{d}}

\begin{document}
\title{Longitudinal dependence of open heavy flavor $\raa$ in relativistic heavy-ion collisions}
\date{\today}

\author{Caio A.~G.~Prado}
\author{Wen-Jing Xing}
\affiliation{Institute of Particle Physics and Key Laboratory of Quark and Lepton Physics (\moe), Central China Normal University, Wuhan, Hubei, \oldstylenums{430079}, China}
\author{Shanshan Cao}
\affiliation{Department of Physics and Astronomy, Wayne State University, Detroit, MI \oldstylenums{48201}, USA}
\affiliation{Cyclotron Institute and Department of Physics and Astronomy, Texas A\&M University, College Station, TX \oldstylenums{77843}, USA}
\author{Guang-You Qin}
\author{Xin-Nian Wang}
\affiliation{Institute of Particle Physics and Key Laboratory of Quark and Lepton Physics (\moe), Central China Normal University, Wuhan, Hubei, \oldstylenums{430079}, China}
\affiliation{Nuclear Science Division, Lawrence Berkeley National Laboratory, Berkeley, CA \oldstylenums{94720}, USA}

\begin{abstract}
  Heavy flavor probes are sensitive to the properties of the quark gluon plasma (\qgp) produced in relativistic heavy-ion collisions.
  A huge amount of effort has been devoted to studying different aspects of the heavy-ion collisions using heavy flavor particles.
  In this work, we study the dynamics of heavy quark transport in the \qgp\ medium using the rapidity dependence of heavy flavor observables.
  We calculate the nuclear modification of $\Bmeson$ and $\Dmeson$ meson spectra as well as spectra of leptons from heavy flavor decays in the rapidity range $[-4.0,4.0]$.
  We use an implementation of the improved Langevin equation with gluon radiation on top of a (3+1)-dimensional relativistic viscous hydrodynamical background for several collision setups.
  We find that the rapidity dependence of the heavy quark modification is determined by the interplay between the smaller size of the medium, which affects the path length of the heavy quarks, and the softer heavy quark initial production spectrum.
  We compare our results with available experimental data and present predictions for open heavy flavor meson $\raa$ at finite rapidity.
\end{abstract}

\maketitle

\section{Introduction}

The quark gluon plasma (\qgp) produced in relativistic heavy-ion collisions is currently the most perfect fluid in nature~\cite{Gyulassy:2004zy, Shuryak:2004cy, Adams:2005dq, Romatschke:2007mq, STAR:2017ckg}.
Tomographic study of the \qgp\ via jet-medium interaction and jet quenching is one of the most important methods for probing such hot and dense nuclear matter \cite{Wang:1991xy,Qin:2015srf,*Qin:2015srf.edited}.
Heavy quarks are particularly valuable probes due to their large masses compared to the \qcd\ scale~\cite{Moore:2004tg, Borsanyi:2010bp,Dong:2019byy}.
Since they are mainly created at the very earliest stages of the collisions, the final state observables from heavy quarks contain cumulative information of the evolution dynamics of the quark gluon plasma.

One of the most common observables pertaining heavy flavor studies is the nuclear modification factor $\raa$, which compares the yields in nucleus-nucleus collisions with proton-proton collisions, giving information on how the \qgp\ interacts with heavy quarks.
In particular, the $\raa$ is important to understand effects that originate from the hot \qcd\ matter produced in the collision and it is commonly associated with parton energy loss through the dense medium.
It is defined as the ratio between the particle spectrum in nuclei collisions $\dd N_{\nuclei}/\dd \pt$, and the spectrum in $\pp$ collisions, $\dd N_{\pp}/\dd \pt$~\cite{Miller:2007ri}:
\begin{equation}\label{eq:raa}
\raa(\pt,y) = \frac{1}{\mathcal{N}}\frac{\dd N_{\nuclei}/\dd \pt \dd y}{\dd N_{\pp}/\dd \pt \dd y},
\end{equation}
in which $\mathcal{N}$ is the average number of binary nucleon-nucleon collisions for a given class of $\nuclei$ collisions.

For the past several years many studies attempted to use $\raa$ to investigate mechanisms of parton transport and energy loss~\cite{Gossiaux:2010yx,He:2011qa,Young:2011ug,Alberico:2011zy,Uphoff:2012gb,Nahrgang:2013saa,Das:2013kea,Song:2015ykw,Cao:2015cba, Cao:2016gvr,Cao:2017hhk,Cao:2017crw,Liu:2017qah,Li:2018izm,Ke:2018tsh,Xing:2019xae}.
The nuclear modification factor is often used together with other observables in order to obtain stricter constraints on phenomenological models.
The anisotropic flow measurements for heavy flavor lead to the so called $\raa\times\vn2$ puzzle~\cite{Das:2017dsh,Scardina:2017ipo,Rapp:2018qla,Cao:2018ews}, since many theoretical model calculation underestimate $\vn2$ though they can describe $\raa$.
Further studies on the flow coefficients and event-by-event fluctuations as well as event shape engineering analysis also rely on $\raa$ for an initial constraint of parameters or validation of theoretical models~\cite{Prado:2016szr,Beraudo:2018tpr,Beraudo:2018bxb,Katz:2019fkc,Katz:2019qwv}.

Most studies so far mainly focus on the mid-rapidity regime.
One may also explore the longitudinal dependence of heavy flavor observables which may put stricter constraints on the currently available models and provide further insight into the dynamics of heavy quarks transport in the \qgp\ medium.
Recent studies along this direction focus on the rapidity dependence of the direct flow of $\Dmeson$ mesons~\cite{Chatterjee:2017ahy,Nasim:2018hyw,Zhang:2019hzn,Adam:2019wnk}, though a limited range of rapidity is used.
We can also investigate the nuclear modification factor which is affected mainly by the path length traversed by heavy quarks through the medium and their initial production spectra.
In the forward rapidity regime, the medium conditions differ from that at mid-rapidity, as the system is smaller and thus the path length is shorter.
In addition, initial heavy quark production spectra in this regime also differ greatly.
Thus, by exploring the behaviour of $\raa$ with respect to rapidity one can obtain the picture of \qgp\ medium in a wider range of phase space.

In this study, we investigate the longitudinal dependence of the $\raa$ of heavy flavor mesons ($\Bmeson$ and $\Dmeson$) as well as electrons and muons decayed from these particles.
We use the three dimensional medium profiles generated from \clvisc\ viscous hydrodynamics code~\cite{Pang:2018zzo} to construct averaged \qgp\ backgrounds for different collision setups.
On top of the hydrodynamics backgrounds, heavy quarks are sampled and allowed to propagate through the medium using a previously developed framework implementing a relativistic Langevin equation with gluon radiation and a hybrid fragmentation plus coalescence model for hadronization~\cite{Cao:2013ita,Cao:2015hia}.
Heavy mesons are allowed to decay into electrons and muons.
Using the analysis framework developed for \dabmod~\cite{Prado:2016szr,Katz:2019fkc} we obtain our final results and compare with currently available experimental data.
Predictions are made for different rapidity bins in the range of $-4.0 < y < 4.0$ on the nuclear modification factor of open heavy flavor mesons.

\section{Elements of the simulation}

In order to simulate the propagation of heavy quarks inside the \qgp\ we use a modified relativistic Langevin equation~\cite{Cao:2013ita,Cao:2015hia} which incorporates two different processes of energy loss inside the medium: quasi-elastic scattering with light partons in the plasma and gluon radiation induced by multiple scatterings.
The Langevin equation can be described by:~\cite{Cao:2013ita,Cao:2015hia}
\begin{equation}\label{eq:CaoLangevin}
    \frac{\dd \vec{p}}{\dd t} = -\eta_D(\vec{p})\vec{p} + \vec{\xi} + \vec{f}_\text{g},
\end{equation}
in which the last term $\vec{f}_\text{g}$ is added to the original Langevin equation and corresponds to the recoil force exerted on the heavy quarks due to the gluon emission.
The other two terms of the equation are the drag force and the thermal force.
Here we assume $\vec{\xi}$ to be independent of the momentum $\vec{p}$ and satisfies the following correlation function,
\begin{eqnarray}
    \big\langle \xi^i (t) \xi^j(t') \big\rangle = \kappa \delta^{ij} \delta(t-t')
\end{eqnarray}
where $\kappa$ is the momentum diffusion coefficient and related to the spatial diffusion coefficient as $D=2T^2/\kappa$.
For all the calculations presented in this work, the spatial diffusion coefficient is set as $D(2\pi T) = 7$
which provides the best description of the experimental data that will be shown later.
The gluon radiation term in Eq.~(\ref{eq:CaoLangevin}) is calculated from the probability of a gluon emission during a fixed time interval, with the gluon emission spectrum given by the higher twist formalism~\cite{Guo:2000nz,Zhang:2003wk,Majumder:2009ge}.
More details on the implementation of this improved Langevin approach can be found in ~\cite{Cao:2013ita,Cao:2015hia}.

The study of longitudinal dependence of observables requires a three dimensional profile of medium evolution.
Therefore, we use the (3+1)-dimensional relativistic hydrodynamics code \clvisc~\cite{Pang:2018zzo}.
In this work we explore three different collision systems: $\AuAu$ at $\snnG{200}$, $\PbPb$ at $\snnT{2.76}$, and $\PbPb$ at $\snnT{5.02}$.
The hydro simulation is initialized with a smooth initial condition using the \trento~\cite{Bernhard:2016tnd} parametrization that mimics the \ipglasma~\cite{Schenke:2012wb,Song:2012kw,Lappi:2006xc} at initial time $\tau_0 = \SI{0.6}{\fm}$ and evolve the medium with an \eos\ described by lattice \qcd\ calculation: \eosPCE, in which the system is partially chemically equilibrated~\cite{Huovinen:2009yb}.
During the evolution of hydrodynamics we set the shear viscosity as $\eta/s = 0.15$, and the system evolves until the freeze-out temperature $\Tfo = \SI{137}{\MeV}$ is reached.
With these setups, the hydrodynamic model is able to provide good descriptions of the soft hadron spectra emitted from the \qgp~\cite{Pang:2018zzo}.

Heavy quarks are initially sampled within the medium before the hydrodynamic evolution.
We determine the initial positions of the heavy quarks production at $\tau = \SI{0}{\fm}$ on the transverse plane using the binary collision distribution obtained from Monte Carlo Glauber model.
The initial momentum distribution of the heavy quarks is calculated using a leading order perturbative \qcd\ calculation~\cite{Combridge:1978kx} including flavor excitation and pair production processes, as well as nuclear shadowing and anti-shadowing effects~\cite{Cao:2015hia,Lai:1999wy,Eskola:2009uj}.
Heavy quarks are allowed to propagate freely in the three dimensional space until $\tau = \tau_0$ when the hydrodynamical evolution begins.
They will then transport through the medium and lose energy according to the Langevin equation until their hadronization occurs at the decoupling temperature $\Td = \SI{165}{\MeV}$.
The hadronization of heavy quarks uses a hybrid fragmentation and coalescence model~\cite{Cao:2013ita,Cao:2015hia}.
Heavy mesons are finally decayed into electrons and muons via \pythia.

\section{Numerical results}

\begin{figure}[t!]
    \includegraphics{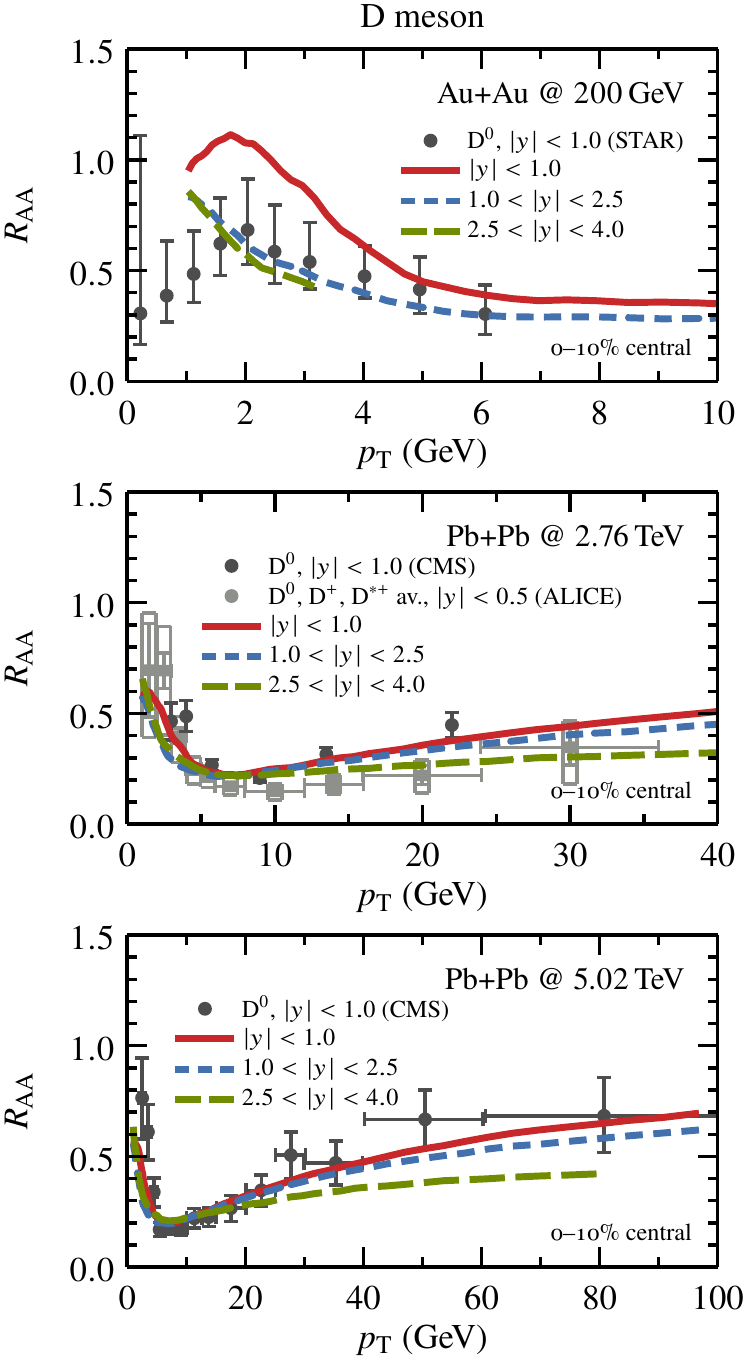}
    \caption{(Color online) Nuclear modification factor of $\Dmeson$ mesons for central collisions in different ranges of rapidity.  Mid-rapidity results are compared with data from \STAR~\cite{Radhakrishnan:2019gbl} (top), \alice~\cite{Adam:2015sza} and \cms~\cite{CMS:2015hca} (middle), and \cms~\cite{Sirunyan:2017xss} (bottom) at their respective collision energies.}
    \label{fig:raaD}
\end{figure}

The results for the $\Dmeson$ meson nuclear modification factor are shown in Fig.~\ref{fig:raaD} for collisions of $\AuAu$ at $\snnG{200}$, $\PbPb$ at $\snnT{2.76}$, and $\PbPb$ at $\snnT{5.02}$.
The solid red curves in the plots correspond to the $\Dmeson$ meson spectra at mid-rapidity which are compared with experimental data.
We observe a good agreement with \cms\ data for both $\PbPb$ collisions throughout the whole $\pt$ range.
Since \cms\ data slightly disagree with \alice, our results overestimate $\raa$ for $\snnT{2.76}$ case in comparison with \alice\ data.
For the lowest energy collision of $\AuAu$ at $\snnG{200}$ our results show consistency with data from the \STAR\ experiment for $\pt \geq \SI{4}{\GeV}$.  At the lower $\pt$ regime a complex interplay of different physical processes is expected to occur.
One important effect is the recombination mechanism which tends to dominate the heavy quark hadronization at this regime.

\begin{figure}[t!]
    \includegraphics{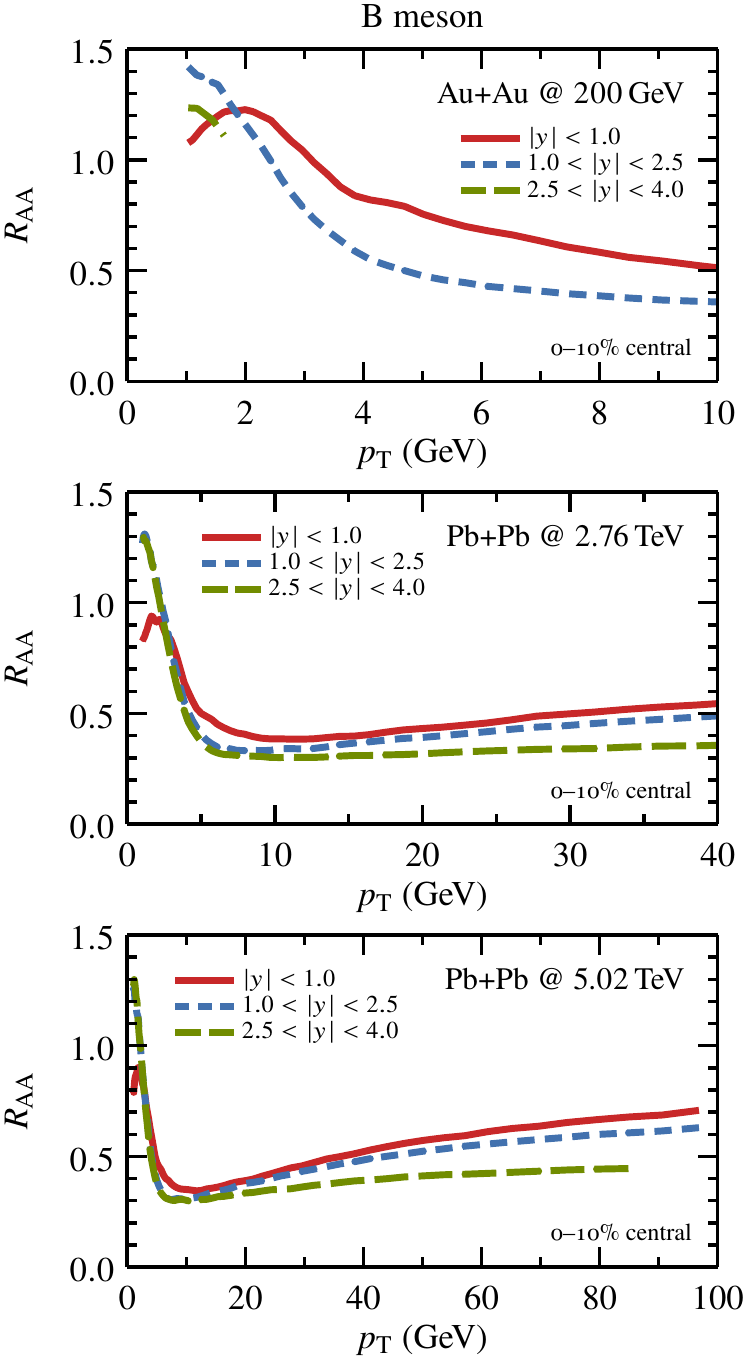}
    \caption{(Color online) Nuclear modification factor of $\Bmeson$ mesons for central collisions in different ranges of rapidity for $\AuAu$ at $\snnG{200}$ (top), $\PbPb$ at $\snnT{2.76}$ (middle) and $\PbPb$ at $\snnT{5.02}$ (bottom).}
    \label{fig:raaB}
\end{figure}

We also show in Fig.~\ref{fig:raaD} predictions for forward rapidity $\raa$ between $1.0 < \abs{y} < 2.5$ and $2.5 < \abs{y} < 4.0$ rapidity ranges.
Lower collision energies reflect in smaller ranges of achievable $\pt$ at large rapidity, as observed in the top panel of the figure for the long-dashed green curve.
When increasing the rapidity, we observe a larger suppression at the high $\pt$ regime, even though the expected medium size in these conditions is smaller.
Since $\raa$ not only depends on the path length experienced by the parton inside the medium, but also on the initial production spectra, we expect these two effects to compete in the final result.
Here, a stronger effect from the initial heavy quark spectra is observed to dominate in this region of $\pt$.

\begin{figure}[t!]
    \includegraphics{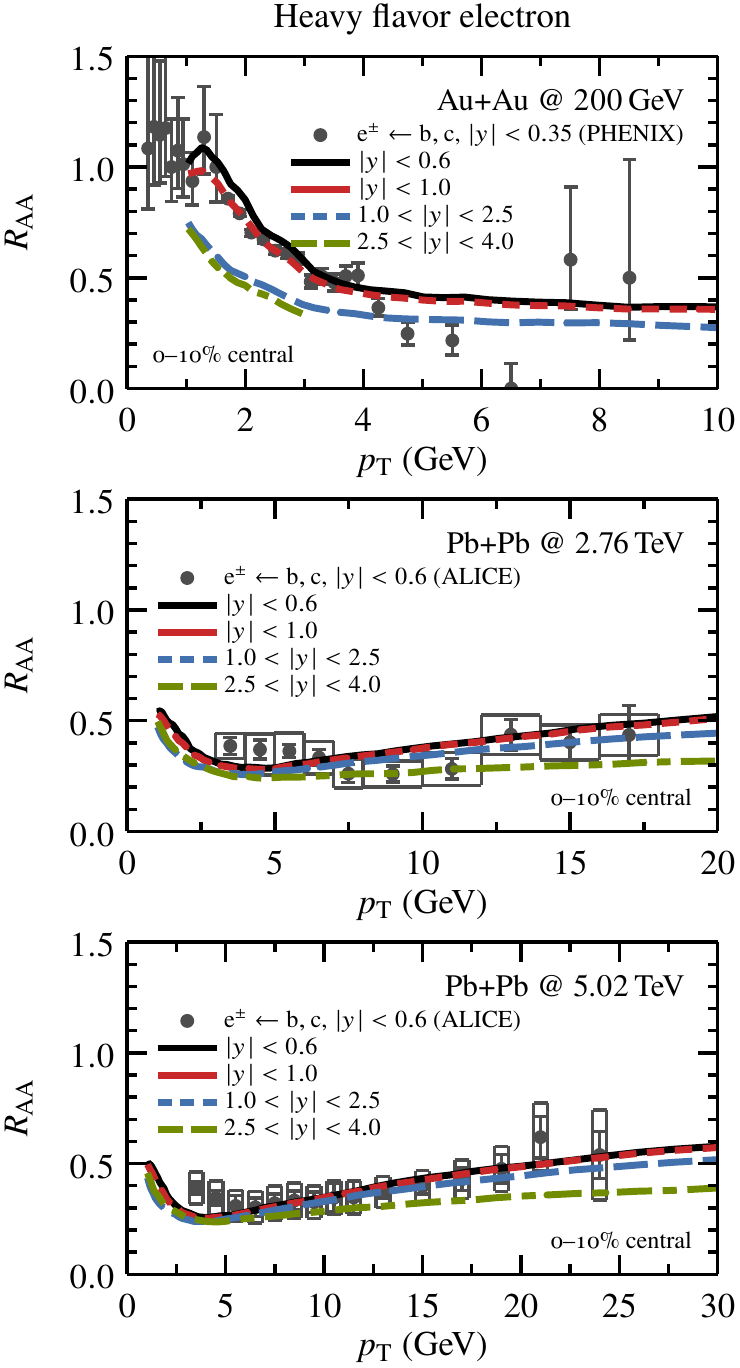}
    \caption{(Color online) Nuclear modification factor of heavy flavor electrons for central collisions in different ranges of rapidity.  Mid-rapidity results are compared with data from \phenix~\cite{Adare:2010de} (top), \alice~\cite{Adam:2016khe} (middle), and \alice~\cite{ALI-PREL-133360} (bottom) at their respective collision energies.}
    \label{fig:raae}
\end{figure}

Using the same simulation conditions as above, we show in Fig.~\ref{fig:raaB} the results for the $\Bmeson$ meson $\raa$.
We observe the same trend as the case for $\Dmeson$ mesons at high $\pt$, where a larger parton suppression is observed for larger rapidity bins.
However, in the case of $\Bmeson$ mesons we see a clearer separation of the curves at low $\pt$, together with a more evident crossing for $\pt < \SI{5}{\GeV}$.
This suggests that the rapidity dependence at low $\pt$ regimes may behave differently than that at high $\pt$.
Our results are consistent with previous observations on the $\pionzero$ nuclear modification factor~\cite{Qin:2007zzf}, though the crossing points seem to be at lower $\pt$ in the case of heavy quarks.

\begin{figure}[t!]
    \includegraphics{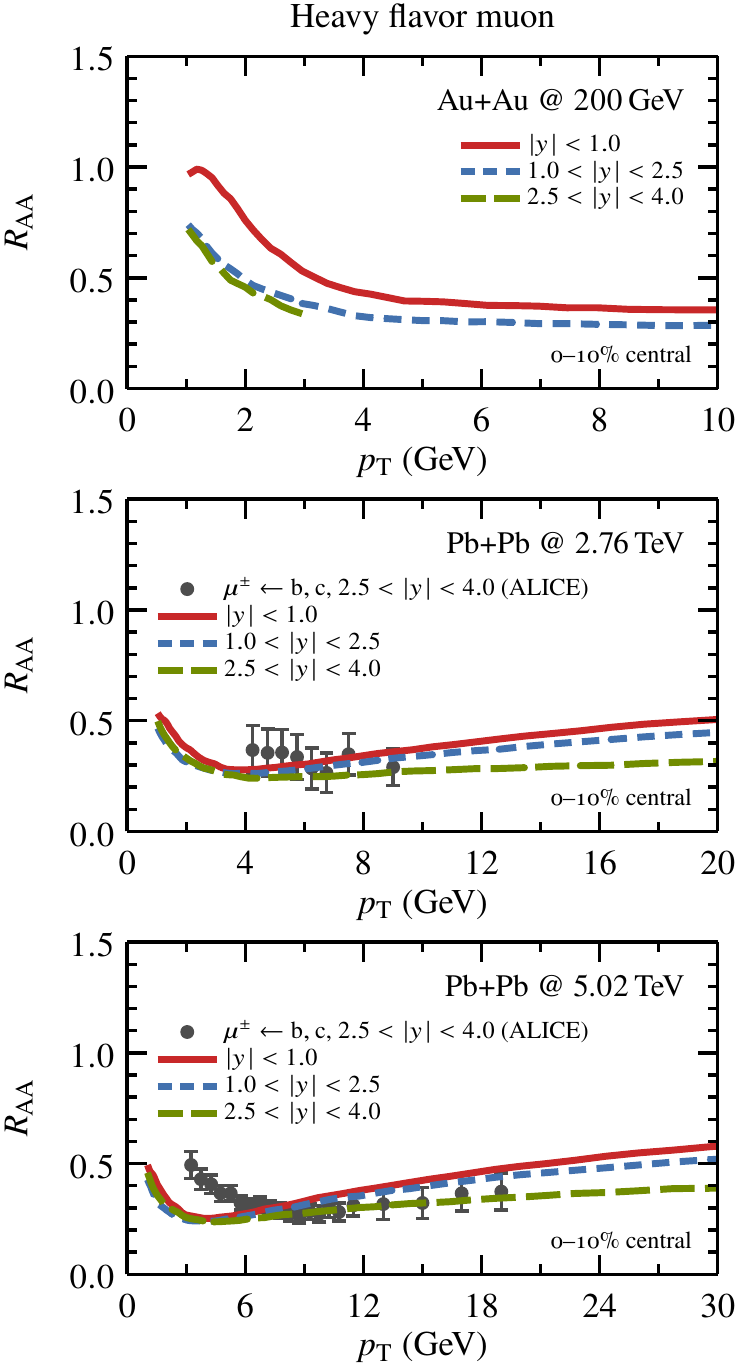}
    \caption{(Color online) Nuclear modification factor of heavy flavor muons for central collisions in different ranges of rapidity for $\AuAu$ at $\snnG{200}$ (top), $\PbPb$ at $\snnT{2.76}$ (middle) and $\PbPb$ at $\snnT{5.02}$ (bottom).  Results are compared with experimental data from \alice~\cite{Abelev:2012qh,ALI-PREL-133394} at forward rapidity.}
    \label{fig:raam}
\end{figure}

After simulation of the heavy meson decays into electrons and muons, we can obtain the nuclear modification factor for these leptons.
In Fig.~\ref{fig:raae} we show heavy flavor electron results.
The plots show an additional curve with a smaller rapidity range of $\abs{y} < 0.6$ to be better compared with data from \alice.
The top plot of the figure shows good agreement with data from the \phenix, even though the rapidity ranges being compared are not exactly the same.
In fact, it is expected that around the mid-rapidity regime, the differences in rapidity have little influence on the results.
We also observe a good agreement within error bars for the $\PbPb$ collisions at $\pt \gtrsim \SI{5}{\GeV}$.
A slight overestimation is observed in the case of $\PbPb$ at $\snnT{2.76}$ which is consistent with the previous result for $\Dmeson$ mesons in comparison with \alice\ in Fig.~\ref{fig:raaD}.
However, overall agreement are observed at both mid and large rapidities.

In addition to the mid-rapidity results, Fig.~\ref{fig:raae} also includes forward rapidity predictions for heavy flavor electrons.
These results reflect what has already been observed for $\Dmeson$ and $\Bmeson$ mesons as we see an increase of suppression for large $\pt$ electrons with large rapidity.
We also note the crossing between spectra with different rapidities at low $\pt$ for $\PbPb$ collisions.

By simulating the heavy flavor mesons to decay into muons, we obtain the results shown in Fig.~\ref{fig:raam}.
We compare our forward rapidity results, showing in long-dashed green curves, with the experimental data for $\PbPb$ collisions at both $\snnT{2.76}$ and $\snnT{5.02}$.
The comparison shows good agreement with data for the larger beam energy at high $\pt$.
In particular, by comparing Figs.~\ref{fig:raam} and~\ref{fig:raae}, we can see that for the high $\pt$ regime there should be enough resolution to study the rapidity dependence of heavy flavor $\raa$ by comparing theory calculations to data, though the current data error bars are still large.
In addition, heavy flavor muon $\raa$ obtained from the simulation falls slightly bellow the experimental data for low $\pt$ as is the case with heavy flavor electrons.
Despite that, for $\snnT{2.76}$ collision, our results are consistent with experimental data within error bars, though at this $\pt$ interval, different rapidity bins are also indistinguishable.
Finally, the nuclear modification factor for $\AuAu$ collisions are predicted in the top panel.

\begin{figure*}[t!]
    \includegraphics{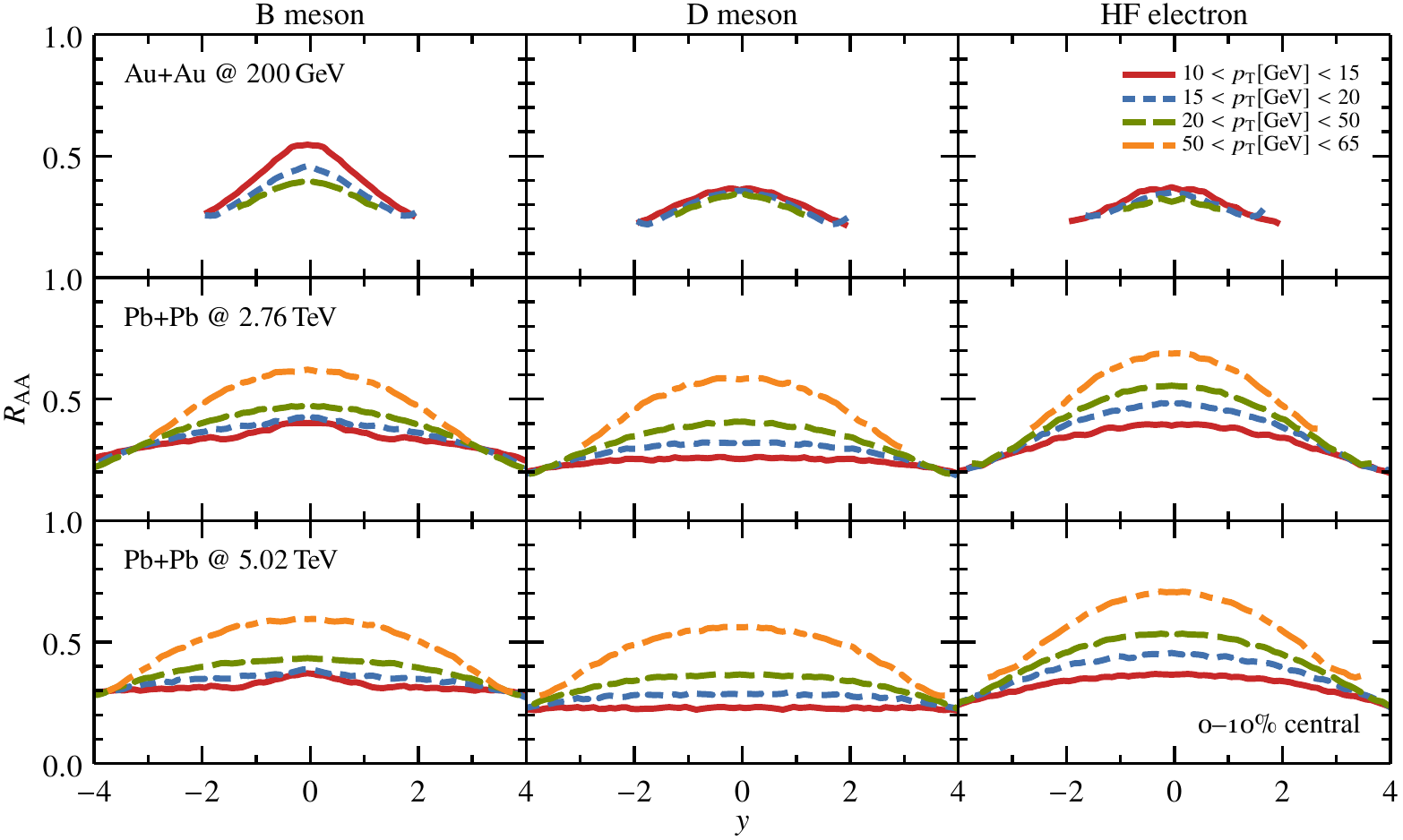}
    \caption{(Color online) Nuclear modification factor for selected $\pt$ ranges differential in rapidity for $\Bmeson$ mesons (left), $\Dmeson$ mesons (middle), and heavy flavor electrons (right) in central collision systems of $\AuAu$ $\snnG{200}$ (top), $\PbPb$ $\snnT{2.76}$ (middle), and $\PbPb$ $\snnT{5.02}$ (bottom).}
    \label{fig:yraa}
\end{figure*}

Fig.~\ref{fig:yraa} shows the rapidity dependence of the nuclear modification factor for different $\pt$ ranges for all the collision systems considered so far.
In these plots, comparing different collision energies shows a widening of the $\raa$ curves with increasing energy.
In other words, the greater the collision energy, the farther from the mid-rapidity regime we observe a deviation on the $\raa$ behaviour.
Another interesting observation is that for two higher energy $\PbPb$ collisions the nuclear modification factor is larger at higher $\pt$, while for $\AuAu$ collisions $\raa$ is smaller for larger $\pt$.
Again this is due to the initial production spectra of heavy quarks: at the same $\pt$ the spectrum is steeper in lower energy nuclear collisions.

The above results suggest that measurements of high-$\pt$ particles at finite rapidity may put more constraints on the $\raa$ for better understanding of heavy flavor transport inside the \qgp.
On the other hand, heavy flavor production in low $\pt$ region can be further tested in the forward rapidity regime and lower collision energies due to its physical complexity.

\section{Conclusions}

In this work we couple the (3+1)-dimensional viscous hydrodynamic medium background modeled by \clvisc\ with a relativistic Langevin equation based transport model incorporating both collisional and radiative energy loss of heavy quarks in order to investigate the longitudinal dependence of heavy flavor nuclear modification factor.
We verified the consistency between our implementation in a 3-dimensional setup and the currently available experimental data at the mid-rapidity regime for different collision energies.
Muon data at finite rapidity were also used to further validate our model.
With our simulation, we provided predictions for forward rapidity $\raa$ of heavy flavor mesons and leptons for three different collision energies.
We find that the smaller size of the medium at larger rapidity and the steeper initial spectra of heavy quarks at larger rapidity compete with each other.
In the end, heavy quarks display small $\raa$ at large rapidity for large $\pt$ regime.
The nuclear modification behavior at low $\pt$ regime is more complex due to the interplay of the recombination and other physics effects.

Further studies on the longitudinal dependence of heavy flavor observables are still necessary, in particular, the dependence of flow coefficients coupled with $\raa$ may provide more sensitive constraints on phenomenological models for better understanding of the quark gluon plasma.
We hope that the predictions presented in this paper encourage the measurement of finite rapidity observables of heavy flavor final state particles with higher precision.

\section*{Acknowledgments}

This work is supported in part by the Natural Science Foundation of China (\nsfc) under Grant Nos. \texttt{11775095}, \texttt{11890711}, \texttt{1861131009} and \texttt{11890714}, by the China Scholarship Council (\CSC) under Grant No. \texttt{201906775042}, by the U.S. Department of Energy (\doe) under grant  Nos. \texttt{DE-AC02-05CH11231} and \texttt{DE-SC0013460}, and by the U.S. Natural Science Foundation (\nsf) under grant No. \texttt{ACI-1550228} within the \jetscape\ Collaboration.

\end{document}